\documentclass[sigconf, screen, nonacm]{acmart}
\usepackage{hyperref}
\usepackage{float}
\usepackage{multirow}
\usepackage{color,colortbl}
\newcolumntype{L}[1]{>{\raggedright\let\newline\\\arraybackslash\hspace{0pt}}m{#1}}
\newcolumntype{R}[1]{>{\raggedleft\let\newline\\\arraybackslash\hspace{0pt}}m{#1}}
\definecolor{Gray}{gray}{0.9}
\AtBeginDocument{%
  \providecommand\BibTeX{{%
    \normalfont B\kern-0.5em{\scshape i\kern-0.25em b}\kern-0.8em\TeX}}}

\setcopyright{acmcopyright}
\copyrightyear{2018}
\acmYear{2018}
\acmDOI{10.1145/1122445.1122456}

\acmConference[Woodstock '18]{Woodstock '18: ACM Symposium on Neural
  Gaze Detection}{June 03--05, 2018}{Woodstock, NY}
\acmBooktitle{Woodstock '18: ACM Symposium on Neural Gaze Detection,
  June 03--05, 2018, Woodstock, NY}
\acmPrice{15.00}
\acmISBN{978-1-4503-XXXX-X/18/06}

\begin{document}

\title{Considerations and Challenges of Measuring Operator Performance in Telepresence and Teleoperation Entailing Mixed Reality Technologies}

\author{Eleftherios Triantafyllidis}
\affiliation{%
  \institution{School of Informatics, The University of Edinburgh}
  \city{Edinburgh}
  \country{United Kingdom}}
\email{eleftherios.triantafyllidis@ed.ac.uk}

\author{Zhibin Li}
\affiliation{%
  \institution{School of Informatics, The University of Edinburgh}
  \city{Edinburgh}
  \country{United Kingdom}}
\email{zhibin.li@ed.ac.uk}


\begin{abstract}
Assessing human performance in robotic scenarios such as those seen in telepresence and teleoperation has always been a challenging task. With the recent spike in mixed reality technologies and the subsequent focus by researchers, new pathways have opened in elucidating human perception and maximising overall immersion. Yet with the multitude of different assessment methods in evaluating operator performance in virtual environments within the field of HCI and HRI, inter-study comparability and transferability are limited. In this short paper, we present a brief overview of existing methods in assessing operator performance including subjective and objective approaches while also attempting to capture future technical challenges and frontiers. The ultimate goal is to assist and pinpoint readers towards potentially important directions with the future hope of providing a unified immersion framework for teleoperation and telepresence by standardizing a set of guidelines and evaluation methods.
\end{abstract}



\maketitle

\section{Introduction}
When we think of controlling a robot or an avatar in a simulation environment, a fundamental question comes to mind, can we truly inhabit that foreign body? The first step towards this vision is to analyse what constitutes embodiment. The homunculus argument would be the best analogy in this case, in which a supposed ``little person”, in this case, an operator, is looking through the person’s eyes and controlling their actions \cite{gregory_2005}. This fallacy would be appropriate in avatarization or tele-embodiment, only that in this case, we would be the ones acting as the ``little person”. 

To better understand embodiment and the underlying factors associated when operating remote avatars or robots, evaluating and modelling human performance is necessary \cite{10.1145/3411763.3443442, 10.3389/frobt.2018.00074, ranscombe_rodda_johnson_2019, 10.1162/PRES_a_00124}. Evaluating user performance in Virtual Environments (VEs) has always been a longstanding goal \cite{doi:10.1177/1541931213601466, 10.1145/3411763.3443442}. With the spike in technological growth and most notably that of Mixed Reality (MR) technologies, the field of telerobotics has advanced rapidly with the most prominent fields of telepresence and teleoperation receiving substantial focus \cite{9076603, 8003431, 8673306}. In these two subfields of robotics, which inherently include humans in the loop, having the means of assessing operator performance on immersive technologies such as those seen in MR remains invaluable \cite{10.1145/3411763.3443442}.

However, with the multitude of different measurement methods that exist to date when measuring operator performance in robotics and VEs, confusion still exists as to which evaluation methods one should use to effectively measure and model user experience \cite{10.1145/3411763.3443442}. This is further aggravated by the lack of a standardized performance metric. Such a metric would help solidify inter-study comparisons among studies, as different manifestations between user studies and the diverse measurement methods that are implemented between current work, limit consistency and transferability of results, ultimately reducing generalization \cite{s15327043hup0504, doi:10.1177/154193129003401712}.

In this short paper, we present the most widely used evaluation methods that are used in the state of the art when measuring human performance in telepresence and teleoperation. We primarily focus on interaction tasks such as pointing, docking and positioning of objects within VEs. Furthermore, we present some challenges and frontiers when attempting to capture the complex and diverse human factors associated with users in MR. Finally, we present some potential strategies to account for these factors and mitigate to some extent, the limitations and complexities associated with modelling user experience. Ultimately, we hope this work will open some important pathways for researchers in the pursuit of a standardized framework or even evaluation method when capturing user experience in VEs \cite{10.3389/frobt.2018.00074}.

\section{Prominent Evaluation Methods in Simulated Telepresence and Teleoperation}
In this section, we briefly present the differences between telepresence and teleoperation while also presenting the most prominent quantitative evaluation approaches used by researchers to evaluate and assess user performance. We focus particularly on human-guided robotics entailing the use of immersive technologies such as those seen in MR technologies. We go over both types of objective and subjective measurement procedures, as both are usually implemented in a study to limit the drawbacks of using either one exclusively \cite{Slater:1994:DPV:2870936.2870938, 1219cba4b03b4d9e81481fd65a53a490}.

\subsection{Telepresence and Teleoperation}
Before we go over the two primary types of evaluating user performance namely objective and subjective responses, we briefly present the differences between telepresence and teleoperation. The field of human-guided robot control is composed of these two major categories. Generally speaking, the difference between telepresence and teleoperation lies primarily in their applications. Telepresence systems are mostly associated with their capability and objective of mimicking the appearance of a person, at a remote location \cite{triantafyllidisrobot, 10.1145/2702123.2702526}. Hence, telepresence systems are primarily associated with communication technologies such as videoconferencing. 

While telepresence systems are very similar to teleoperation systems in allowing a person to be ``present'' in a remote physical location, teleoperated systems are mostly associated with applications concerning manipulation tasks \cite{triantafyllidisrobot, 9000554}. These can range from remote surgery, space exploration, defence applications, handling of dangerous materials and search and rescue missions \cite{sheridan_1995, 10.5555/645443.653103, 10.1016/S1084-8045(03)00017-1, ALVAREZ2001395}.

Consequently, the primary benefit of human-guided robotics is seen in scenarios and tasks that would otherwise be too expensive, difficult and most importantly too dangerous for humans \cite{10.1007/s10514-017-9677-2}. As both types of systems aim to provide humans with the capability of remotely controlling a robotic body; maximising body-ownership, overall immersion, situational awareness and ultimately elucidating human perception remains a longstanding goal \cite{triantafyllidisrobot, 9076603, 10.1145/3411763.3443442}. The first step towards this goal is for researchers to agree on a set of standardized metrics for assessing human performance on emerging technologies (e.g. VR, AR and MR), methods and applications.

\subsection{Measuring Objective Responses}
The most popular objective measurement techniques used in telepresence and teleoperation in current literature remain spatial-related and time-based metrics \cite{9076603, 10.1145/2702123.2702526, triantafyllidisrobot}. In addition to these, some studies also include behavioural and physical measures such as Electroencephalograms (EEGs), Electrocardiograms (ECGs) and even Functional Magnetic Resonance Imaging (fMRIs) \cite{8211636, 10.3389/fnins.2020.00040, doi:10.1177/1729881419888042, 1641190, DBLP:books/cu/L2020}.

However, measuring distance error, accuracy and the required time to complete tasks, movements and interactions in a virtual environment entailing MR technologies are still the dominant methods of evaluating user experiences, particularly due to the low cost of implementation \cite{8003431, 10.1145/3385956.3422092, 9076603, 8797975, doi:10.1080/10447318.2015.1039909, triantafyllidisrobot, doi:10.1177/0278364919842925, 10.1145/2702123.2702526}.

Most studies use a mixture of the aforementioned objective responses, but ultimately the use of a standardized objective metric for measuring performance is missing \cite{9076603, 8003431, 8673306, doi:10.1177/0278364919842925}. This severely limits inter-study comparability and transferability \cite{10.1145/3411763.3443442}. However, this can be mitigated to some extent if one looks at Fitts' Law \cite{fitts1964information, fitts1954information}. Fitts' Law is one of the most widely used performance metrics attempting to capture human movement in the field of Human-Computer Interaction (HCI) \cite{10.1207/s15327051hci0701_3, doi:10.1080/00140139108967324, 10.1145/1753846.1753867}. The formulation predicts the Movement Time (MT) of how long it takes for users to point to a target on a screen. More specifically, given a generic mouse pointer, the law predicts how much time it would take for a user to point to a target location, given the target's distance \(A\) and target's width \(W\). The law is formulated as:
\begin{equation}
\begin{gathered}
MT = a + b \cdot ID \text{ , } ID = log_2 \left ( \frac{2A}{W} \right ),
\end{gathered}
\label{eqn:fitts_law_original}
\end{equation}
where the logarithmic ratio between \(A\) and \(W\) produces the so-called Index of Difficulty (ID) which quantifies the task difficulty and is measured in bits/sec. As we can infer, this formulation quantifies the difficulty of a task based on spatial data and predicts human movement in a time-based approach. Hence, the law can effectively combine both spatial and time-based measurements under one formulation. Another advantage of the law is that it can be applied in interactions with 2D/3D user interfaces and also manipulation tasks (e.g. including gravity, friction and etc.), thus applicable to both telepresence and teleoperation respectively. 

However, movements in MR in particular, are so diverse and spatially complex, as it entails moving and rotating in full 3D space, that simply using Fitts' law has its limitations. While originally developed for 1D translational tasks, the law has been extended to 2D \cite{welford1968fundamentals, 10.1207/s15327051hci0701_3, doi:10.1080/00140139508925153} and even to some extend to 3D space \cite{CHA2013350, murata2001extending}. Nonetheless, a higher dimensional model of Fitts' Law in 3D space is still missing \cite{10.1145/3411763.3443442}. This is particularly attributed to the complexities associated with such dimensions, the degrees of freedom entailed in VEs and translational as well as rotational variations \cite{10.1145/3411763.3443442, 8998368, doi:10.1177/0018720810366560}. 

Nevertheless, it would be more helpful and likely leading to more reliable observations, if researchers would focus on either Fitts' Law or for that matter any well-established objective tool or formulation, instead of relying on a multitude of different time-based and spatial related measurement tools. For example, different studies assessing user experiences may include a diverse use of different objective metrics. Ultimately, this would result in the lack of consistency of the chosen metrics among these studies, aggravating future comparability and conclusiveness by researchers.

\subsection{Measuring Subjective Responses}
While objective metrics are almost always the chosen tools for assessing operator performance in VEs, including subjective responses are fundamental when attempting to assess user experience and increase generalization \cite{doi:10.1177/154193128402801103, 10.1145/2702123.2702526, 8673306, 9076603, 8211636}. The NASA-Task Load Index, also known as NASA-TLX, is one of the most prominent approaches to measuring the subjective cognitive demand operators spend on completing a set of tasks in telerobotics \cite{hart_staveland_1988}. The questionnaire follows a 7-point Likert scale approach entailing the assessment of mental demand, physical demand, temporal demand, effort, performance and even frustration. Another useful questionnaire used in teleoperation assessment is the System Usability Scale (SUS) \cite{brooke1996sus}. Following a 5-point Likert scale, it measures the overall system usability of a system. 

Both of the aforementioned questionnaires have been used extensively in assessing emerging technologies, such as MR, within the context of teleoperation and telepresence \cite{8673306, 10.1145/3385956.3422092, 9076603, 8211636, doi:10.1177/0278364919842925}. More commonly used in telepresence rather than in teleoperation scenarios, the NASA Situation Awareness Rating Technique (SART) is another 7-point Likert scale questionnaire, fairly extensively used in measuring the perceived feelings of presence in remote environments \cite{10.1145/2702123.2702526, doi:10.1518/001872095779049499}. Both the NASA SART and TLX share similarities in terms of scales and inter-correlation between the concepts of situational awareness and workload demand respectively \cite{doi:10.1518/107118191786755706}.

As with their objective counterpart, consistency among subjective responses is limited in existing work. This is primarily attributed to certain user studies creating their own questionnaire which not only introduces very specific questions to measure user experience but also the Likert scale responses vary significantly \cite{doi:10.1177/0278364919842925, 10.1145/2702123.2702526, SCHNACK201940, 8673306}. For example, finding the same custom questionnaire in a study attempting to assess the overall user experience of new emerging technologies, is highly unlikely due to both the variations of introduced questions and the number of points in the Likert scale. The reason for studies introducing such questionnaires is the lack of a standardized subjective method of assessing user / operator experience within the domain of teleoperation and telepresence. In the end, comparing the results of either the NASA-TLX, SUS or any other well-established questionnaire among different studies, is likely to result in more useful and reliable observations than a multitude of different and diverse types of questionnaires.

\section{Evaluation Challenges and Frontiers}
To this point, we can infer that perhaps the most challenging factor researchers are faced with when evaluating human performance is the multitude of the different measurement methods that can be used. User studies that attempt to capture such information, usually lack consistency when it comes to the selection of such methods which severely aggravates inter-study comparability, verification and transferability. However, there are still some additional challenges when attempting to assess user experience in VEs entailing MR technologies, some of which are detailed below.

\subsection{Depth Accommodation and Visual Perception in MR}
Vision is supported by numerous studies to be our dominant sense \cite{MCINTIRE201418, Rock594}. One study even quantified vision to be contributing to about 70\% of the total sensory system \cite{Heilig1992ELCD}. Hence, an important aspect to take into account when evaluating user performance in VEs, especially with the use of stereoscopic vision as associated with MR, are the effects of depth perception \cite{10.1145/3290605.3300437, 9076603}. It is widely shown that users overestimate their ability to perceive depth distances in VEs when trying to reach for a target \cite{7164348, doi:10.1177/154193129503902006, Witmer:1998:JPT:1246749.1246755}. While some mitigation strategies exist to limit distance overestimation and degraded longitudinal control, such as increasing the resolution of displays \cite{8798026, kenyon2014vision, 10.1080/0014013032000121624}, it is still important to further research this area. This may lead to a framework or an assessment tool in overcoming these depth limitations and increase our understanding of the underlying factors of our visual-motor perception.

\subsection{Human Related Factors}
Immersion and user assessment in MR is still a complex phenomenon as it is highly influenced by human factors not limited to one's own personality, age, health, previous exposure to technologies and even cognitive ability \cite{10.1145/1228716.1228753, doi:10.1177/2041669515593028, 10.1167/iovs.02-0361, 10.1145/3029798.3038369}. Furthermore, being under the influence of stimulants (e.g. coffee or alcohol), tiredness and concentration can also affect all of the possible evaluation approaches one would choose \cite{10.1145/1753846.1753867}. While modelling and accounting for these in existing evaluation approaches may prove to be particularly challenging, consistency among studies can be accomplished during participant recruitment \cite{10.1145/3411763.3443442}. More specifically, to mitigate to some extent for the aforementioned human-related factors, researchers could account for these prior to commencing experiments and retaining consistency among participant recruitment. For example including or excluding participants based on certain factors in a study (e.g. handedness, visual acuity etc.) and comparing these findings between other studies retaining the same consistency, would likely result in more reliable conclusions and observations than without. It would thus be beneficial for user studies to clearly state the state of the participants and the specific selection criteria used.

\subsection{Body Ownership, Embodiment and Multi-Modal Interfaces}
As we can infer to this point, operator/user embodiment is a fundamental part of the overall user experience in scenarios of remote robotic control and even avatarization  \cite{ranscombe_rodda_johnson_2019}. If there was an interface, that would accommodate all the necessary sensory modalities and stimulate all the human senses in such a way that it would feel ``real”, we would be able to experience ``true” embodiment. While we are still far from achieving this, with the rise of immersive emerging technologies (e.g. VR, AR and MR) we are slowly progressing towards this goal \cite{DBLP:books/cu/L2020}. When embodying an avatar in VEs or a remote robotic system, namely tele-embodiment, the user and the foreign body are physically detached from each other. This constitutes the overall experience unnatural and can lead to decreased levels of overall immersion and situational awareness \cite{9076603}. A potential solution to overcome the decreased levels of embodiment is the use of multimodal interfaces, which aim to provide solutions to increase the feeling of body ownership to the further extent of increasing overall human performance \cite{yanco_drury_2004, 1308838, Jennett:2008:MDE:1393652.1393920}. For example, increased tactile perception can directly increase user performance by increasing embodiment \cite{10.3389/frobt.2018.00074}. One work even exhaustively studied the effects of visual, haptic and auditory feedback and concluded that manipulation tasks entailing picking and placing of virtual objects, is positively affected by bi-modal and even more so by tri-modal sensory feedback \cite{9076603}. Perhaps, with further research directed towards multimodality and MR technology, we might be able to use robots to experience our surrogate selves one day.

\section{Conclusion}
\balance
In this work, we presented a brief overview of the most widely used quantitative methods in measuring user / operator performance in VEs associated with telepresence and teleoperation. Furthermore, we identified certain challenges and frontiers researchers are likely going to be faced with when attempting to derive a set of standardized assessment tools. The derivation of such sets of standardized methods would not only increase comparability and transferability of results among different manifestations of user studies but likely increase the overall generalization of the observations by researchers. We hope that through this work, we have aided readers in a brief overview of existing evaluation methods used to date and the potentially important factors that would need to be taken into account towards the endeavour of providing the means of standardizing a set of evaluation tools when capturing user experience.

\section{Acknowledgements}
This research is supported by the EPSRC CDT in Robotics and Autonomous Systems (EP/L016834/1).

\bibliographystyle{ACM-Reference-Format}
\bibliography{references}
\end{document}